\documentclass[aps,twocolumn]{revtex4}

\usepackage[final]{graphics}
\usepackage{amssymb}
\usepackage{amsfonts}
\usepackage{epsfig}

\begin{document}




\title{Dynamic critical behavior of the classical anisotropic BCC Heisenberg antiferromagnet}




\author{Shan-Ho Tsai$^{a,b}$, Alex Bunker$^a$, and D. P. Landau$^a$}
\affiliation{
$^a$ Center for Simulational Physics, University of Georgia, Athens, GA 30602\\
$^b$ Enterprise Information Technology Services, University of Georgia, Athens, GA 30602}

\begin{abstract}
Using a recently implemented integration method [Krech {\it et. al.}] based 
on an iterative second-order Suzuki-Trotter decomposition scheme, we have 
performed spin dynamics simulations to study the critical dynamics of the 
BCC Heisenberg antiferromagnet with uniaxial anisotropy. This technique allowed
us to probe the narrow asymptotic critical region of the model and estimate 
the dynamic critical exponent $z=2.25\pm 0.08$. Comparisons with 
competing theories and experimental results are presented.

\end{abstract}


\maketitle
\section{Introduction}

Despite their relative simplicity, classical Heisenberg models can be used 
to describe the dynamic critical behavior of magnetic materials whose ions 
have large effective spin values. For example, RbMnF$_3$ is
well modeled by an isotropic Heisenberg antiferromagnet, whereas MnF$_2$ and 
FeF$_2$ are good physical realizations of the Heisenberg antiferromagnet with
weak and strong uniaxial anisotropies, respectively.

The effect of uniaxial anisotropy on the critical dynamics has been the subject
of many theoretical \cite{dynscal,modecpg,rngt} and experimental studies 
\cite{fef2,mnf2}. Although both indicate that the 
dynamic structure factor has a dominant purely diffusive longitudinal 
component and a suppressed propagative transverse component, there is still
controversy about the dynamic critical exponent $z$. While a dynamic scaling 
argument \cite{dynscal} and mode-coupling theory \cite{modecpg} indicate 
that $z=2$, a combination of renormalization group theory and 
$\epsilon$-expansion \cite{rngt} predicts that 
$z=2+\alpha/\nu$. Using the Ising value $\nu=0.6289(8)$ from a high 
resolution Monte Carlo 
study \cite{FL} and $\alpha=0.110(5)$ from field theory \cite{GZJ}, the latter 
prediction is $z=2.175(10)$. Experiments on FeF$_2$ \cite{fef2} have not had
enough resolution to distinguish between these two predictions, and on MnF$_2$
\cite{mnf2}
the exponent obtained was that of the isotropic antiferromagnet, $z=1.5$. Presumably this very low value is due
to crossover effects resulting from the weak anisotropy in MnF$_2$.

Our approach to this problem is to use spin dynamics simulations with 
an efficient integration scheme. The dynamic critical exponent is estimated
using a dynamic finite-size scaling theory and compared with the predictions
by the different theories and with experiments.

\section{Model and Methods}

The Hamiltonian of the model is given by
\begin{equation}
\mathcal{H}=J\sum_{<{\bf r},{\bf r}'>}{\bf S_r}\cdot {\bf S_{r'}} - 
D\sum_{\bf r}({S_{\bf r}}^z)^2
\end{equation}
where ${\bf S_r}=(S_{\bf r}^x,S_{\bf r}^y,S_{\bf r}^z)$ is a 
three-dimensional classical spin of unit length, 
$<{\bf r},{\bf r}'>$ denotes nearest-neighbor pairs of spins coupled by 
the antiferromagnetic exchange interaction $J>0$, and $D$ is the uniaxial 
single-site anisotropy. We use $D=0.0591J$ and the inverse critical temperature
$1/T_c=0.478k_B/J$ appropriate for MnF$_2$ \cite{LBC}.
We consider body-centered-cubic lattices with linear sizes $12\le L\le 60$, 
corresponding to $2L^3$ sites, and periodic boundary conditions. 

The time evolution of the spins is governed by coupled equations of motion, 
and the dynamic structure factor $S({\bf q},\omega)$, for momentum transfer
${\bf q}$ and frequency $\omega$, observable in neutron scattering
experiments, is the Fourier transform of the space-displaced, time-displaced 
spin-spin correlation function $C^k({\bf r} - {\bf r'},t)$, defined as
$$
C^k({\bf r} - {\bf r'},t) =\langle S_{{\bf r}}^k(t)S_{{\bf r'}}^k(0)\rangle-\langle S_{{\bf r}}^k(t)\rangle\langle S_{{\bf r'}}^k(0)\rangle.
$$
with $k=x, y,$ or $z$. Because of the periodic boundary conditions, we can
only access a set of discrete values of momentum transfer given by 
$q=2\pi n_q/L$, where $n_q=\pm 1,\pm 2,..., \pm L/2$.

Equilibrium configurations at $T_c$ were generated using a hybrid Monte Carlo 
method where each hybrid step consisted of two Metropolis 
sweeps through the lattice and eight Wolff cluster spin flips. Typically 3000 
hybrid Monte Carlo steps were used to generate each equilibrium configuration, 
which was then used as an initial spin configuration in the integration of 
the coupled equations of motion. We use $2040$ initial configurations to 
compute the averages. For the largest lattice used ($L=60$), there are 
432,000 coupled equations to integrate. 
These numerical integrations were performed to a maximum time of 
$t_{max}=400/J$, using an iterative second-order Suzuki-Trotter 
decomposition method implemented by Krech {\it et al} \cite{krech}, with 
two iterations and a time step of $\delta t=0.04/J$. 
This method consists of performing explicit rotations of a spin about its 
local effective field. The lattice is divided into two sublattices 
$\mathcal{A}$ and $\mathcal{B}$, and a formal solution of the equations of 
motion yields 
$y(t+\delta t)=e^{({\rm A + B})\delta t}y(t)$, where $y=(y_A,y_B)$ denotes
collectively all the spins, and the matrices A and B are the infinitesimal 
generators of rotation of the spin configurations $y_A$ on sublattice 
$\mathcal{A}$ at fixed $y_B$ and of the spin configurations $y_B$ on 
$\mathcal{B}$ at fixed $y_A$, respectively. The spin configurations on 
sublattices $\mathcal{A}$ and $\mathcal{B}$ are updated in succession, 
using the second-order Suzuki-Trotter 
decomposition of the exponential operator \cite{SU}, given by 
$e^{({\rm A + B})\delta t}=e^{{\rm A}\delta t/2}e^{{\rm B}\delta t}
e^{{\rm A}\delta t/2} + O(\delta t^3)$. A similar scheme using a 
fourth-order decomposition of the 
exponential operator has also been tested in Ref.\cite{krech}. The iterative
procedure is necessary because of the single-site anisotropy\cite{krech}.

The finite integration time introduces oscillations when the Fourier transform
of $C^k({\bf r} - {\bf r'},t)$ is taken to produce $S({\bf q},\omega)$ 
and we smooth them out by convoluting the spin-spin
correlation function with a Gaussian resolution function in frequency, with
characteristic width $\delta_{\omega}$. The dynamic structure factor thus
obtained is denoted as $\bar S^k({\bf q},\omega)$.

The dynamic critical exponent $z$ can be determined from 
dynamic finite-size relations \cite{kun}, given by
 ${\omega\bar S_L^k({\bf q},\omega)}/{{\bar{\chi}_L}^k({\bf q})}=
G(\omega L^z,qL,\delta_{\omega}L^z)$
 and 
\begin{equation}
\bar\omega_m=L^{-z}\bar\Omega(qL,\delta_{\omega}L^z),
\label{omegam}
\end{equation}
where ${\bar{\chi}_L}^k({\bf q})=\int_{-\infty}^{\infty}\bar S_L^k({\bf q},\omega){d\omega}/{2\pi}$ is the momentum dependent susceptibility
 and $\bar\omega_m$ is a characteristic frequency, given by
 $\int_{-\bar\omega_m}^{\bar\omega_m}\bar S_L^k({\bf q},\omega){d\omega}/{2\pi}$ $={\bar{\chi}_L}^k({\bf q})/2.$
To estimate $z$ we choose the width of the resolution function to be 
$\delta_{\omega}=0.015(60/L)^z$
so that the function $\bar\Omega(qL,\delta_{\omega}L^z)$ in Eq. (\ref{omegam})
 is a constant if $qL$ is fixed, yielding
\begin{equation}
\bar\omega_m\sim L^{-z}.
\label{omegamLz}
\end{equation}
Because $\delta_{\omega}$ depends on $z$, this exponent had to be determined 
iteratively. The coefficient $0.015$ of $\delta_{\omega}$ was chosen 
empirically as a compromise between effectively reducing the oscillations in
$S({\bf q},\omega)$ and not excessively broadening its structure.

\section{Results and Discussion}
\begin{figure}[ht]
\centering
\leavevmode
\includegraphics[clip,angle=0,width=7.3cm]{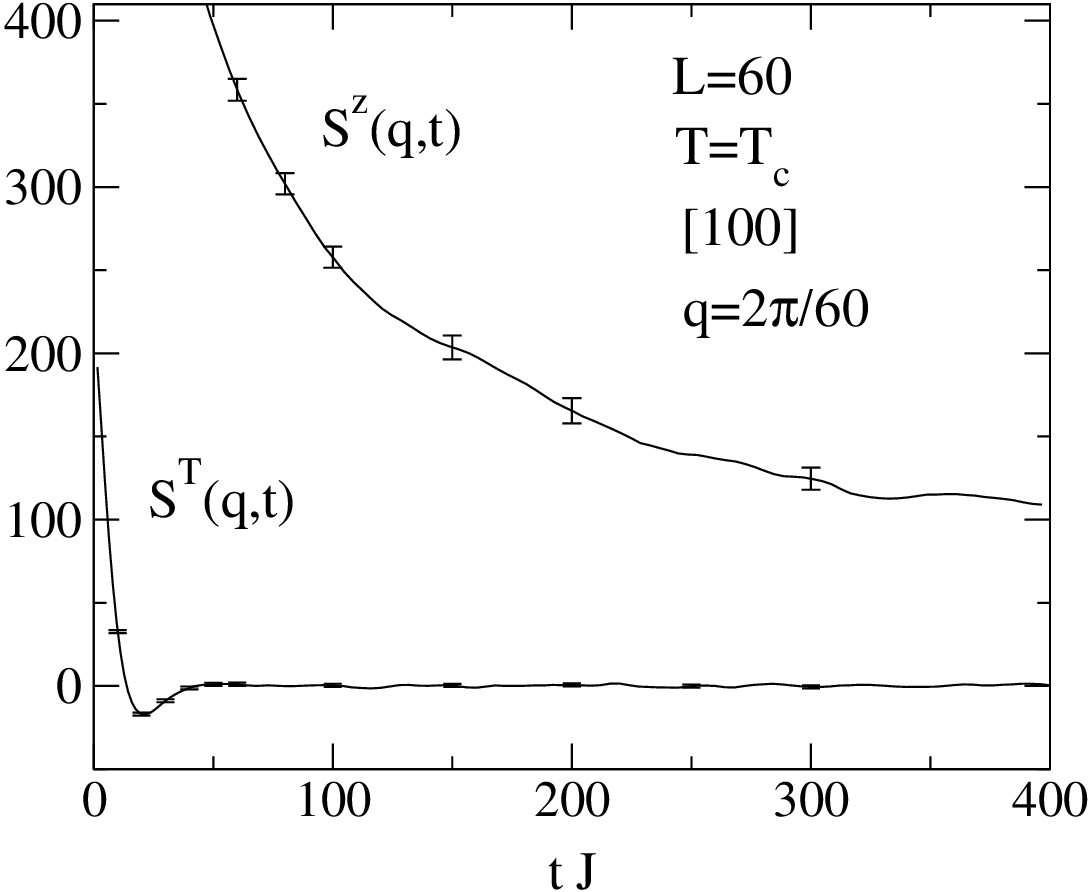}
\caption{Longitudinal and transverse components of the space-Fourier-transformed correlation function. A few characteristic error bars are shown.}
\label{sqt}
\end{figure}

Our simulations, performed at the critical temperature for MnF$_2$ \cite{LBC},
 show that while the transverse component $S^T({\bf q},t)$ of 
the space-Fourier-transformed correlation function $S({\bf q},t)$ has a short
relaxation time, the longitudinal component $S^z({\bf q},t)$ decays extremely
slowly [see Fig.(\ref{sqt})], and this requires that the equations of motion
be integrated to very long times. Although our $t_{max}$  
is still not large enough for the longitudinal component to 
decay to a value that is close to zero, it is a significant improvement over
the maximum time we could reach with more standard integration methods, 
such as the fourth-order predictor-corrector method, with our current 
computing resources. Whenever not shown, error bars are smaller than the 
symbol sizes in the figures.

\begin{figure}
\centering
\leavevmode
\includegraphics[clip,angle=0,width=7.5cm]{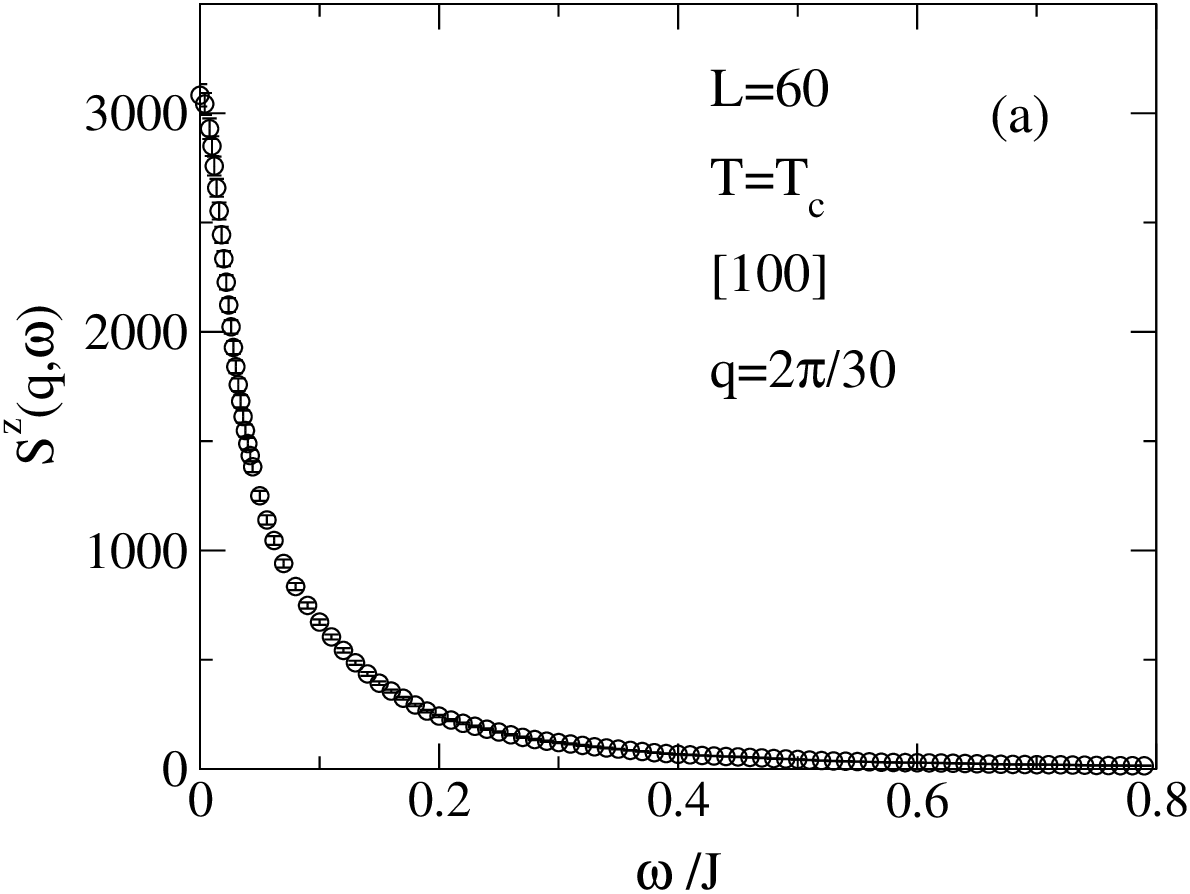}
\includegraphics[clip,angle=0,width=7.5cm]{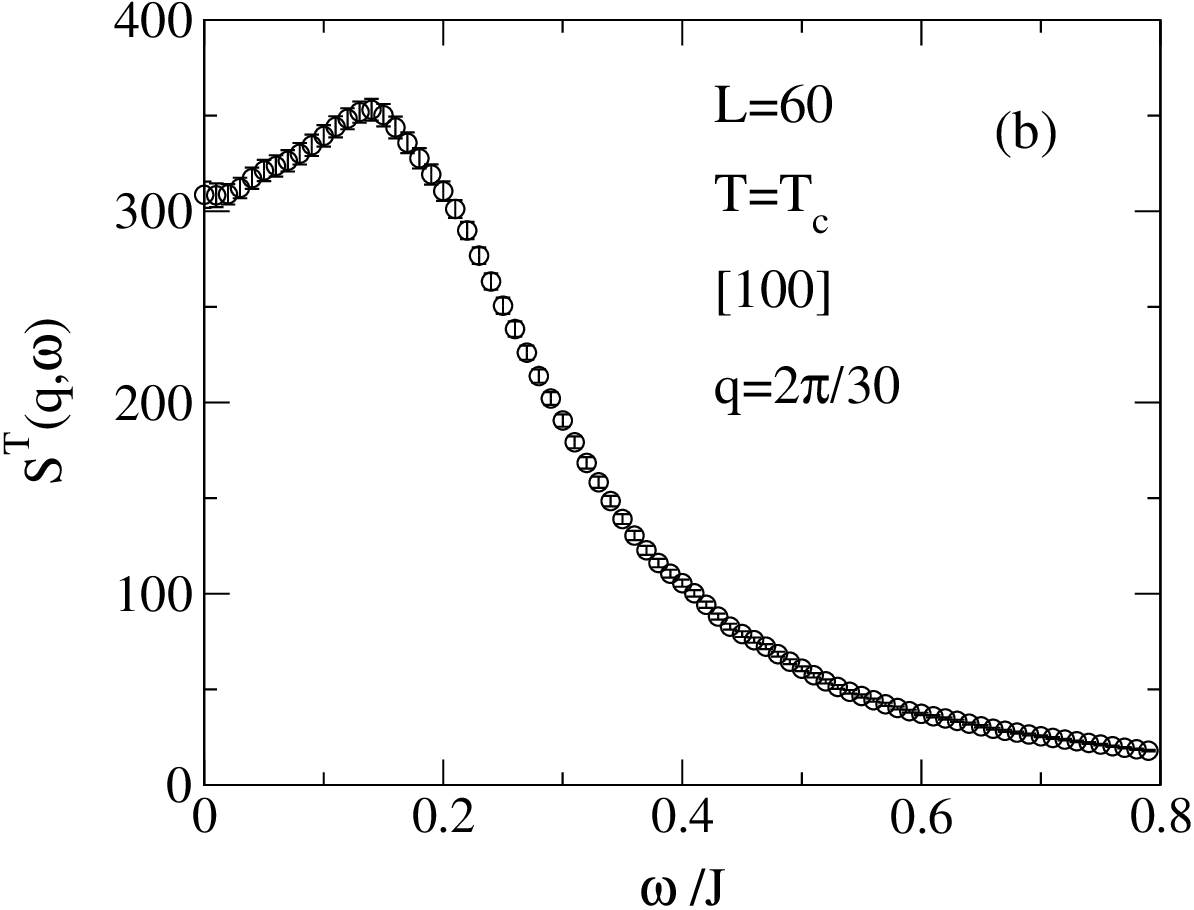}
\caption{(a) Longitudinal and (b) transverse components of the dynamic 
structure factor, with $\delta_{\omega}=0.015$.}
\label{sqw}
\end{figure}

The longitudinal component of the dynamic structure factor, 
$S^z({\bf q},\omega)$, shown in Fig.(\ref{sqw}a), has a purely dissipative
behavior, as indicated by the single central peak. In contrast, the 
transverse component $S^T({\bf q},\omega)$ contains a propagative mode, 
indicated by the spin wave peak in Fig.(\ref{sqw}b), with a possible 
small central peak as well. 
A comparison between Figs.(\ref{sqw}a) and (\ref{sqw}b) 
shows that the longitudinal component has a much larger intensity. 
\begin{figure}
\centering
\leavevmode
\includegraphics[clip,angle=0,width=7.0cm]{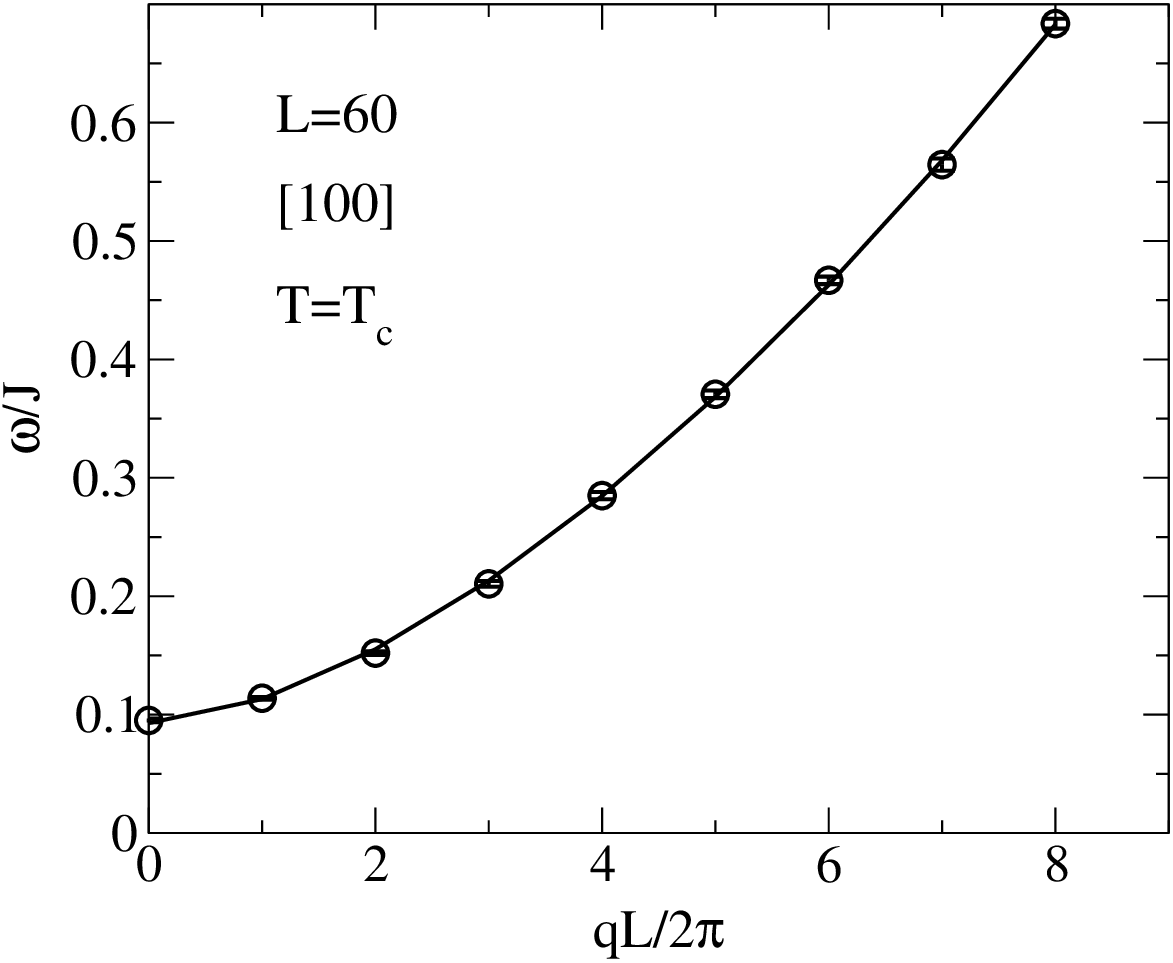}
\caption{Dispersion relation for the spin wave excitation in 
$S^T({\bf q},\omega)$ for small values of $q$.}
\label{disp}
\end{figure}
\begin{figure}
\centering
\leavevmode
\includegraphics[clip,angle=0,width=7.5cm]{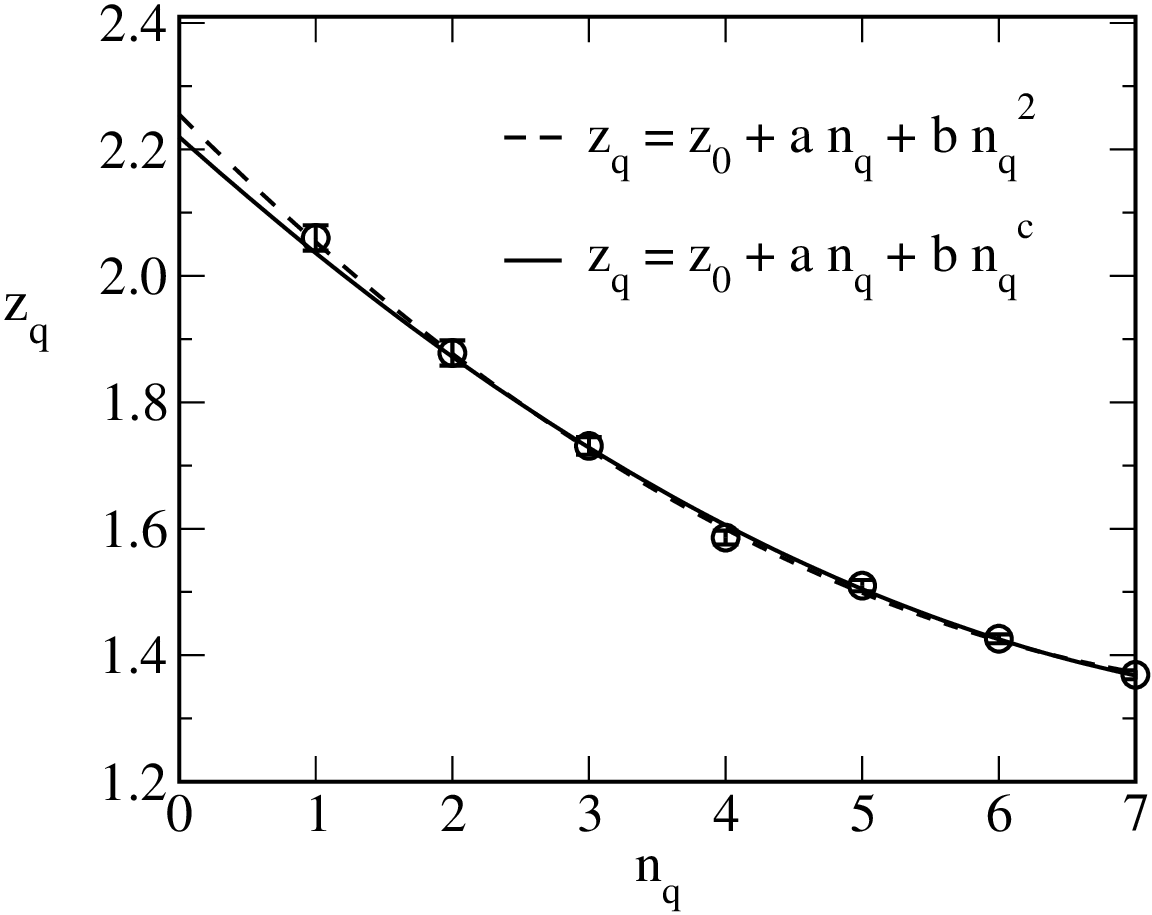}
\caption{Estimate of the dynamic critical exponent for different $n_q$. 
Analysis done with $L=30,36,48$ and $60$.}
\label{zqnq}
\end{figure}
The dispersion curve $\omega(q)$ of the spin waves along the [100]
direction is shown in Fig.(\ref{disp}), where the solid line represents
a fit to the function $\omega=\omega_0+cq^x$. The non-zero value of the
frequency in the limit as $q\to 0$ is characteristic of an anisotropic
model. These results for the dynamic structure factor are in qualitative
agreement with theory and experimental results. 

An estimate for the dynamic critical exponent $z$ was obtained iteratively
using Eq.(\ref{omegamLz}) for different values of $n_q$. 
Such estimates are denoted as $z_q$ and are shown in Fig.(\ref{zqnq}). 
For the weakly 
anisotropic system considered here, the onset of the asymptotic critical 
region occurs at very low values of $q$, accessible only with very large 
lattice sizes (and not accessible to experiments yet). 
Therefore, a better estimate for $z$ is given by $z_0$,
obtained by extrapolating $z_q$ to the limit $q\to 0$. We fitted $z_q$ with
 the function $z_q=z_0 + a{n_q}+b{n_q}^c$, using $n_q=1,2,...,{n_q}^{\rm max}$,
where $z_0$, $a$, $b$, and $c$ are fitting parameters. To check the 
robustness of the extrapolated $z_0$, we have also performed fittings with
 fixed $c=2$. The systematic change in the value of $z_0$ for both types of 
fittings was studied as the number of points included in the 
fittings increased, from ${n_q}^{\rm max}=4$ and $5$, for the fittings with
$c=2$ and variable $c$, respectively, to ${n_q}^{\rm max}=15$. For  
${n_q}^{\rm max}$ up to 7, the $\chi^2$ of the fittings is reasonable, and 
there is no clear systematic change in the value of $z_0$; hence, the 
average $z_0$ from these fittings yields $z=2.25\pm 0.08$, which is the first
numerical estimate of the dynamic critical exponent for this model.  

\section{Summary and Conclusions}
Because of difficulties for 
experiments to probe the critical region, experimental data have not yet 
been able to distinguish between competing theories. While limited by 
finite lattice size and finite integration time, simulations offer the 
hope of shedding light on the differences between theories and experiment. 
Although not yet conclusive, our estimate of $z=2.25\pm 0.08$ is slightly
larger than, but consistent with, the prediction 
by the renormalization group theory. It is not consistent with
 mode-coupling theory and the dynamic scaling prediction.

\vspace{0.2cm}

This work was partially supported by NSF grant DMR-0094422.
Simulations were performed on the Cray T90 and IBM SP (Blue Horizon) at SDSC.

\end{document}